# BLCS: Brain-Like based Distributed Control Security in Cyber Physical Systems

Hui Yang[1, *], Kaixuan Zhan[1], Michel Kadoch[2], Yongshen Liang[1], Mohamed Cheriet[2]

1: Beijing University of Posts and Telecommunications, Beijing, 100876, China.
2: École de technologie supérieure, Montreal, Canada.
*Corresponding author e-mail: yanghui@bupt.edu.cn

**Abstract:** Cyber-physical system (CPS) has operated, controlled and coordinated the physical systems integrated by a computing and communication core applied in industry 4.0. To accommodate CPS services, fog radio and optical networks (F-RON) has become an important supporting physical cyber infrastructure taking advantage of both the inherent ubiquity of wireless technology and the large capacity of optical networks. However, cyber security is the biggest issue in CPS scenario as there is a tradeoff between security control and privacy exposure in F-RON. To deal with this issue, we propose a brain-like based distributed control security (BLCS) architecture for F-RON in CPS, by introducing a brain-like security (BLS) scheme. BLCS can accomplish the secure cross-domain control among tripartite controllers verification in the scenario of decentralized F-RON for distributed computing and communications, which has no need to disclose the private information of each domain against cyber-attacks. BLS utilizes parts of information to perform control identification through relation network and deep learning of behavior library. The functional modules of BLCS architecture are illustrated including various controllers and brain-like knowledge base. The interworking procedures in distributed control security modes based on BLS are described. The overall feasibility and efficiency of architecture are experimentally verified on the software defined network testbed in terms of average mistrust rate, path provisioning latency, packet loss probability and blocking probability. The emulation results are obtained and dissected based on the testbed.

## I. Introduction

Along with the rapid evolution of IoT and industrial applications, developing 5G systems and then evolving into beyond 5G (B5G) have become an inevitable trend which interconnects hundreds of terminals from physical environment with large bandwidth to address the burgeoning machine-to-machine communications. In such scenario, as the bridge between physical world and cyber world, the cyber-physical system (CPS) has operated, controlled and coordinated the physical systems integrated by a computing and communication core, which is successful applied in internet industry or industry 4.0 [1-3]. In order to accommodate CPS services, the radio and optical networks (RON) has become one of the most important supporting physical cyber infrastructures carrying for CPS which can take advantage of both the inherent ubiquity of wireless technology and the large capacity of optical networks [4]. RON can provide the low latency and high bandwidth communication and rapid calculation for industry system control and orchestration [5].

Not only considering the accessibility of the network, the cyber security has also drawn much attention from academia and industry for CPS in all walks of life [6], especially against the increasing cyber-attacks and their sophisticated behaviors. The centralized control approach in cyber usually encounter with the threats of collapse and information leakage disastrously, resulting in poor scalability, low reliability and high response time. [7]. The distributed fog-RON (F-RON) for multiple domain can avoid or reduce the risk compared to centralization which safeguards the privacy of each domain, while much lower delay communication should be addressed in F-RON [8]. However, as disposing of the unified controller, distributed architecture must face the trusted issue among multiple controllers with the information of their own domains. Once one of the controllers is malicious intruded by cyber-attacks, the privacy information of each other controller may be exposed for distributed computing and communications [9]. Therefore, there is a tradeoff between security control and privacy exposure in F-RON for CPS operation. For all we know, trusted control security against cyber-attacks in distributed way hasn't been addressed so far in cyber layer of CPS without private exposure.

On the other hand, artificial intelligent (AI) technology can deal with an amount of big data that come from different sources of information to protect the CPS from zero-day cyber-attacks and predict the future attackers' misbehavior [10]. However, few researches consider the security issue if the controller has been broken through by attacks. Recently, brain-like computation and communication [11, 12] is a promising technology in artificial intelligence area as a viable solution with associative recalling the relationship like brain's central nervous system using multiple knowledge bases. Therefore, it is significant to apply brain-like technology to strengthen the distributed cyber security in F-RON for CPS.

The efficient service accommodation of F-RON based on machine learning has been presented in our previous work [13]. In this paper, we propose a novel brain-like based distributed control security (BLCS) architecture for F-RON in cyber physical systems, by introducing of a brain-like security (BLS) scheme. Instead of disclosing the private information of each domain, BLCS can accomplish the secure cross-domain control among tripartite controllers' verification in the scenario of decentralized F-RON for distributed computing and communications. BLS utilizes parts of information to perform control identification through deep learning of behavior library. The overall feasibility and efficiency of architecture with BLS are experimentally verified on the software defined network testbed. The rest of the article is organized as follows. Functional modules of BLCS architecture are illustrated including various controllers and CPS platform. Then we discuss the interworking procedures for control security in BLCS. BLS scheme is presented based on the architecture. The testbed is established and analyzed with numerical results.

## II. Security Challenge in CPS

In CPS system, distributed controllers have private data for the entire network, such as topology information. If one of the controllers is compromised, the attacker can obtain the private data of CPS system and then use the data to destroy the normal operation of the entire system. In this case, the privacy leakage problem aroused is extremely serious. Traditional ways of sharing private data provide multiple opportunities for malicious attacker. The zero-day attack provides a clear sample of the attacks.

To construct trusted connection, the controller cannot complete mutual trust with no private data. Providing all the private data to the authentication will cause a security leak once it is infected. Providing partial information is a promising solution. Security authentication and route provisioning in case of privacy protection are challenging in CPS system due to the difficulties in using partial information for security authentication. Therefore, new technologies are needed to provide intelligent control over CPS for consistent and effective security authentication as well as route provisioning.

## III. Brain-Like based Distributed Control Security Architecture

The traditional pattern of network control is approved in centralized cyber center. Though unified control has the characteristics of maintainability, responsiveness and convergence, in case of malicious attacks, the centralized control center will face the risk of collapse disastrously. Additionally, the centralized center is difficult to prevent the intentional deception with peer-peer relationships in distributed CPS networks. On the other hand, the distributed network also may expose the private information of other domain for the routing and path provisioning convergence [14]. Note that, there is a tradeoff between security control and privacy exposure in F-RON for CPS management. Therefore, we present brain-like based distributed control security architecture to implement distributed trusty and security for cyber control and industry service provisioning in CPS, which is shown in Fig. 1. Instead of unified management in network, the proposed architecture removes central control function towards distributed domain, which splits up the attacked risk and reduces control latency and connection cost to enhance the CPS's service responsiveness. Without revealing all the personal information, BLCS architecture can perform the distributed cross-domain control using brain-like computing which can associate the relationship among partly information with knowledge bases to achieve the trust routing and control.

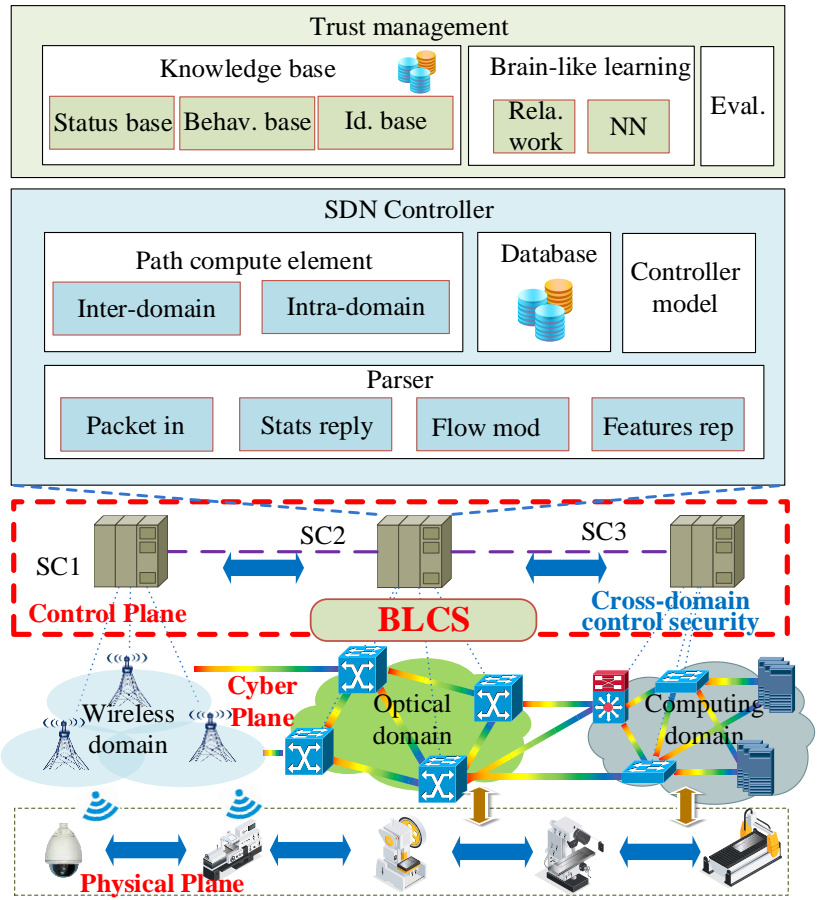

Fig. 1 The architecture of brain-like based distributed control security in F-RON for CPS.

*A. Network Architecture*

BLCS architecture in F-RON for CPS is illustrated in Fig. 1. The physical entities such as robotic arm and camera are interconnected with distributed multi-domain fog wireless and optical network, which is deployed radio, optical and fog computing resources respectively and called as cyber plane. Here, every domains are software defined by the controllers in a centralized way. Aiming to realize the distributed control security, the secure controller in BLCS are interworked with SDN controllers (SC) to constitute a distributed control plane deployed with brain-like computing in order that each industrial production service in CPS can be accommodated with secure cross-domain mixed communication path efficiently. In such control plane, brain-like security scheme is addressed with relation network among partial information to prevent deliberate deception based on the security controller. Each controller (e.g., SC1) only manages its own optical domain network (e.g., domain 1) so as to safeguard its private information. Meanwhile, according to relation network built by brain-like learning, each controller maintains the partial virtual topology and routing behavior of other domain (e.g., domain 3) to verify that the executed computation results whether to correct in SC (e.g., SC3). The motivations for secure control has been driven by two forces. Firstly, the BLCS adopts brain-like computing mechanism in CPS in order to perform the secure anonymous distributed control without private information disclosure addressed by relation network speculation, which promotes the cyber control security with high privacy protection and saves the network cost effectively. Secondly, once relation network has been integrated, the associate relationship will be much clearer using brain-like computing to enhance the network security with the increasing of the services, which accommodate the secure CPS service efficiently with rapid response and global optimization of optical, wireless and calculation resources.

*B. Functional model for BLCS*

In order to implement the secure distributed control, the SDN controllers are used to cooperate with each other and secure controller for trusted cross-domain path provisioning with wireless, optical and computing resources in F-RON. Note that the controllers are extended to support BLCS and setup the functional entities are illustrated in Fig. 1.

In SDN controller, cyber information of its own domain is gathered in physical information module, while other domain's partly virtual topology and routing behavior of should be sustained in virtual information module to be primed for route verification. The controller model is the core engine of SDN controller such NOX and opendaylight to orchestrate and manage the F-RON network in a unified manner. Cross-domain path calculation is executed in PCE base on an integrated graph of F-RON networks with secure constraints, which including inter-domain and intra-domain computation with brain-like credible identification in secure controller. Parser performs continuous radio and optical spectrum allocation and decides the routing path using OpenFlow, which contains the messages of packet in, stats reply, flow modification and features reply. It can control the spectrum bandwidth and center frequency elastically for CPS services, through setting all the related antennas and optical switches on the calculated path or network.

The trust management is a vital module in the architecture shown in Fig.1. Trust management mechanism enables the controller capable to sustain trust and mitigate the risk of communication and information sharing with malicious controller. The trust management model consists of three modules including the knowledge base (KB) module, brain-like learning module, and evaluation module. The KB is updated periodically based on the statistics about the state collected from the physical cyber. The BLS module builds the identification-status-behavior relational network (ISB-RN) of knowledge bases, realizing trust calculation based on partial information. The evaluation module calculates the controller's honesty, reliability, and collaboration through the ISB-RN so as to determine whether a controller is trusted to execute cyber operation.

The core functional entity is secure controller in trust management model. In secure controller, the information of network status such as virtual topology, routing behavior and controller identification should be collected as knowledge bases for brain-like computing and verification. The brain-like learning module can address the secure controller access control and computation authentication. Initially, all the controllers are identified with distributed consensus. They can be in a distributed secure network way as the foundation of credible route. The relation network module verifies the trusty and security of inter-domain path calculation among other controller so as to validate the results dependable with the learning from knowledge of partly information. The neural network (NN) can perform the deep learning to help the relation network establish.

## IV. Cooperation Procedure for Brain-Like Cross-Domain Security

In CPS, the SCs are automatically working and cooperating with each other to execute a certain task assigned from physical entities. However, the SC is vulnerable to be attacked, so that a malicious SC can gain access to private data in the communication channels. Processing and storage of the untrustworthy SC result in unpredictable loss or changes to data, thereby compromising other controllers. Therefore, a controller should be properly identified and validated to be confirmed whether the controller is an appropriate correspondent before starting the interaction with it. In our scheme, the SC only needs to send partial information that does not involve private information to the target controller for trusted authentication and secure routing. Based on the partial information, the destination controller associates with the meaning represented by this information, thereby verifying the identity of the sender. This protects the sender from transmitting its private information to the cyber.

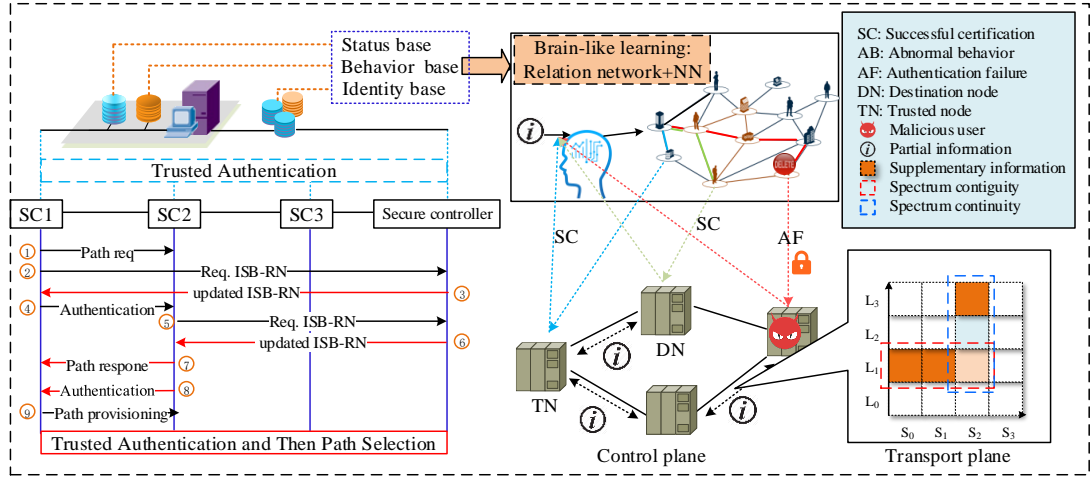

Fig. 2 Illustration of brain-like learning scheme.

Fig. 2 shows how the scheme works. The controllers are divided into two roles including the leading secure controller and the receiver. As the leader of multiple controllers, the secure controller periodically synchronizes the network information to realize the consistency among multiple controllers. In addition, mutex lock is used to ensure synchronization when updating security controller and distributed controller.

The scheme is applied to the control plane and transport plane of optical networks shown in Fig.2. Here, the transport plane is tightly coupled to the physical layer of the optical network, which includes spectrum contiguity and spectrum continuity constraints. The integrated information can be supplemented based on partial information under two constraints of optical network. New spectrum information will build the new relation with existing partial information in knowledge base. Then, the relational network of established spectrum information can be used for security certification. The specific process of security authentication is as follows: SC2 receives the authentication request and the partial information (i.e., the situation of occupied spectrum) of SC1. It can be determined that the slots with orange color are occupied due to the relation of spectrum contiguity and spectrum continuity in S-RN. SC2 can confirm whether the candidate resources from request have been occupied already and then judge whether SC1 is trusted.

In the control plane, the first step is to provide knowledge bases input for the relation network (RN) model, and the RN builds the identification-status-behavior relational network (ISB-RN) between knowledge bases. There is a general undetermined interaction in the knowledge bases. For instance, the processes interact universally such as control and calculation. Interactions follow different types of relations in different spaces. To obtain the complex relations, RN must be able to reason with the existing databases and then learn to infer unknown relations. Thus, we can provide identification, status and behavior description directly into RN. We use tags to mark these inputs which indicate their relative position in the support set, and process this information independently using the same Long Short-Term Memory (LSTM). We notice that such setting invokes minimal prior knowledge like a particular identity will be associated with a state or an operation. After processing the input, in the multi-layer perception (MLP) of the RN, the layers take relational reasoning to be the process of understanding the ways in which information are connected using such understanding to accomplish the construction of the ISB-RN. In fact, these layers provide an architecture on which the model can learn to partition information and learn to calculate the interaction between the partitioned information. For instance, the calculation unit calculates the angle at which the motor rotates, while the motor rotates to control the robotic arm. Here, the robotic arm adjusts the posture and completes the automobile manufacturing. So calculation unit, motor and automobile manufacturing will build a relationship among them.

After the above process, the SC waiting for verification sends part of the information to the destination SC to verify whether it is a trusted object. We process the partial information with LSTM in the way of the first step. Then, recalling relations of ISB-RN in associative fashion to confirm whether to trust. The relationship lookup is first performed, and the relationship between the partial information sent by the controller and each location in the ISB-RN is compared. The relationship recursive search is performed in the ISB-RN until the relationship of the partial information is found. Then, the authentication of partial information can be verified with associative recall. According to associative recall, we confirm it as a trusted SC if there is no exception. Finally, the untrusted SCs will be detected, while we will use the result for secure routing. When one physical object sends a request to establish a trusted route to the optical domain and the computing domain, it can avoid untrusted SCs and use the greedy algorithms for routing safely in trusted domains.

In the proposed model, we illustrate the complex relationships required for trust and routing provisioning across domains of the CPS infrastructure and services. Establishment of ISB-RN and trusted routing authentication is explained with an example.

Example: $I_{identify}=\{i_1, i_2, …, i_n\}$ denotes the identification base, $B_{behavior}=\{b_1, b_2, …, b_n\}$ is the behavior base, and $S=\{s_1, s_2, …, s_n\}$ represents the status base. Each trusted information is defined as combination of $I_{identif}$, $B_{behavior}$ and $S_{status}$, which are constrain of ISB-RN. We assume that SC1, SC2 are trusted controllers, and SC3 is a malicious SC. When SC1 needs to verify its legal identity to SC3. It has to send partial non-sensitive information to SC3, which not only protects the identity of SC1 but also performs trusted authentication. And based on this part of the information, SC3 make an associative recall judgment whether SC1 is credible. The flow steps in the distributed control security and trusted routing authentication process between SCs shown in Fig. 2 is elaborated as below:

Create ISB-RN of knowledge bases: In light of knowledge bases, the RN model constructs the ISB-RN as shown in Fig. 3. The relation instance is the result of relational reasoning in RN. For example, $i_1$ is associated with $i_2$, $i_3$, $s_1$ and $b_2$, just like, under the status of robot $i_2$ and conveyor $i_3$, a motor $i_1$ receives the feedback of the computing domain after the request, and adjusts the speed action $b_2$ in state $s_1$. When new affairs occur, the ISB-RN will update periodically.

Send partial information: The verifier sends a small portion of the information that is not private but sufficient for verification. For instance, the partial information may be part of event information, part of spatial-temporal information, even part of attribute information. Here, we take event information as an example. We can define event information $I_{event}=\{P_{id}, t, l_{loc}, a_{type}\}$ in CPS, where $P_{id}$ is the physical event identifier, $t$, $l$, $a_{type}$ are the occurrence time, location and affairs type, respectively, the event information included in knowledge base suggestively describes whether the SC is trusted. We assume SC1 sent the information $I_{sc1}=\{t, l_{loc}, a_{type}\}$ related its identification to make SCs of another domain trust it.

Carry out associative recall: Recursive association search is performed in ISB-RN in light of received partial information $I_{sc1}$ for trusted verification. When SC3 received $I_{sc1}$, SC3 processes this information. First, according to $I_{sc1}$, event $I_{event}$ is recalled, then the handler of the event is associated, and finally, the SC1 is confirmed the credible. At the same way, SC3 sent its information $I_{sc3}$ to SC1 to verify whether SC1 can trust SC3. When SC1 receives the information from SC3, in the ISB-RN, some abnormal behaviors of SC3 are retrieved, which may cause error handling of the sender information. Therefore, SC3 is judged to be the malicious controller. Hence, SC1 repeats the process with SC2 and find SC2 trusted.

Create the trusted routing connection: When it is verified the trust, create a trusted connection between SC1 and SC2. Once locating and avoiding malicious nodes, we use the greedy algorithm to calculate routes in the trusted SC domain as the final routing result, further improving routing security.

## V. Brain-like Security for CPS

For the first time, we design a cross-domain distributed security scheme based on brain-like learning. The model foundation is based on the RN. The design goal behind BLS is to constrain the functional form of the neural network such that it can capture the core common attributes of relations and enable the formation of associative data structures. It is built upon the RN model and able to store presented relation patterns and recall missing patterns in an associative brain-like fashion.

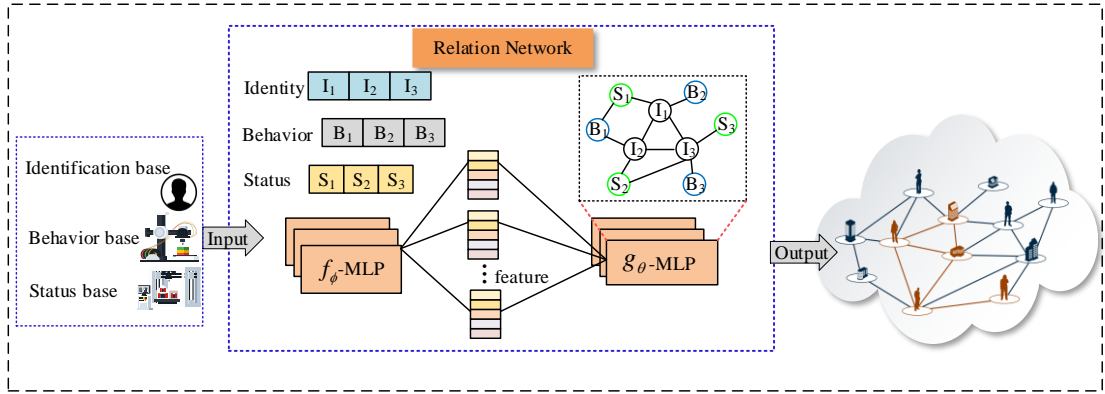

Fig. 3 The architecture of relation network model.

*A. Relation Network*

Relation extraction is prerequisite for BLS. Therefore, we first need to design the RN model with a structure prepared for information relational reasoning. The RN consists of two parts as shown in Fig. 3: feature extraction MLP $f_\emptyset$ and feature relation mapping MLP $g_\theta$. The RN architecture has the capacity to compute relations, just as CNNs have the capacity to reason about spatial and the recurrent neural networks have the capacity to reason about sequential dependencies. We applied the RN to information authentication control which hinges on relational reasoning. The learning model differs from the traditional neural network in the way that it can train the model without labeling, address the evolution in network structure, and infer the indirect relationship between the data. To ensure that the relationship is correctly extracted, we utilize multi existing knowledge bases to heuristically align with information.

The simplest form of the RN model is the following composite function. In Equation (1), the functional form indicates that BLS considers the underlying relationship between knowledge bases. It means that BLS does not necessarily know which relation actually exists, nor the actual implication of any specific relation. Hence, BLS has to learn to reason the existence and implication of relations.

$$\text{BLS}(R) = f_\emptyset(\sum_{i,j} g_\theta\left(i_m, b_p, s_k\right)) \qquad (1)$$

Here the input is sets of identification base $I=\{i_1, i_2, i_m..., i_n\}$, behavior base $B=\{b_1, b_2, b_p..., b_n\}$, and status base $S=\{s_1, s_2, s_k..., s_n\}$. In Equation (1), $f_\emptyset$ and $g_\theta$ are MLPs with parameters $\emptyset$ and $\theta$, respectively, and the parameters are learnable synaptic weights such that RN can be end to end differentiable. We also use LSTM with 32 units to process *I, S, B* as input. The output of $g_\theta$ is referred to as a "relation"; therefore, the mission of RN is to infer the ways in which $i_m$, $b_p$, $s_k$ are relational. RN uses $g_\theta$ to compute relations of information. This can be deemed to as a single function operating on a batch of $i_m$, $b_p$, $s_k$ pairs, where each batch is a particular *Identification-Behavior-Status* pair.

*B. Associative Recall*

Retrieving all information from a small amount of information according to ISB-RN in an associative brain-like fashion, thereby preventing the risk of privacy information leakage [15]. A two-layer memory network is constructed to realize our goal, including the input layer and the relation memory layer. The input layer is used to input relation vector of the RN. The relation memory layer is used as storage of information from the input layer, relation vector information can be stored in this layer.

In the relation memory layer, different sub-networks are used to store the different sets of relation groups, and neural nodes are used to indicate the distribution of data's relation. The input relations are stored in the weight of such nodes. The relation vector with a relation group label is input into the system. According to the label, the model addresses the corresponding sub-network in the relation memory layer, and incrementally learns to add the new input to corresponding sub-network. If the input relation vector does not pertain to any group existing in the relation memory layer, the new input vector will be turned into the first node of a new relation group which will be added to the relation memory layer.

For retrieving all relations, the relation memory is supposed to recall relation vector stored in model so that incomplete information input can be recognized. When presenting partial information, if the relation of the partial information is available, then we will find the corresponding node in the memory layer. If the information's relation is unavailable, the k-nearest neighbor rule will be used to determine which relation vector belongs to. At test time, one is given partial information, which is used to iteratively address and read from the memory looking for relational information to retrieve all relations. At each step, the collected information from the memory is cumulatively added to the original query to build the context for the next round. At the last iteration, the final retrieved information and the most recent query are combined as the final result.

## VI. Performance Analysis and Results Discussion

The efficiency of the proposed BLCS architecture is evaluated on our SDN testbed, which consists of 30 physical objects, wireless domain, optical domain and computing domain, which communicate with distributed controllers to access various service. This scenario is emulated and established on our testbed which has a multi-core server with 8 physical 2.20GHz CPU cores and 2 NVIDIA GTX TITAN XP GPU cores, and the code is based on TensorFlow 1.13.1 in Ubuntu 18.04.2. To train the proposed model, routing trusted authentication requests from physical objects with discrete event simulation are considered for accessing services. All these routing trusted authentication requests along with knowledge bases are given as input to the proposed RN model for security path provisioning. The RN model is run for 2000 epochs using the feature set mentioned above. Once the security route is decided, the proposed scheme is evaluated on the basis of performance metrics such as average mistrust rate, path provisioning latency, packet loss probability, and blocking probability.

Impact on Average Mistrust Rate: To prove the robustness of our algorithm, the average mistrust rate is emulated under the condition of a high proportion of malicious controllers. Fig. 4(a) shows that BLCS can maintain a low average mistrust rate, even though a collusion situation occurs in such case. The variations in average mistrust rate with respect to the ratio of malicious controllers are shown in Fig. 4(a). As seen in this figure, the scheme without BLCS is much higher than the proposed scheme, and it rises faster as the ratio of malicious controller increases. This is because establishing a connection directly without performing trusted authentication increases the probability of connecting to a malicious controller, as opposed to the proposed scheme.

Impact on Path Provisioning Latency: Fig. 4(b) shows the impact on path provisioning latency with increasing traffic load. As can be seen from the figure, the proposed scheme decreases the path provisioning latency. This is because the continuously updated ISB-RN becomes clearer as the network load increases, and the trusted authentication and path provisioning can be performed more quickly. This means that the controller can directly determine which specific domain to authenticate and route in line with ISB-RN when processing the request. It is not blindly interacting with other controllers and maintain trust with highly reliable controllers, thus reducing path provisioning latency.

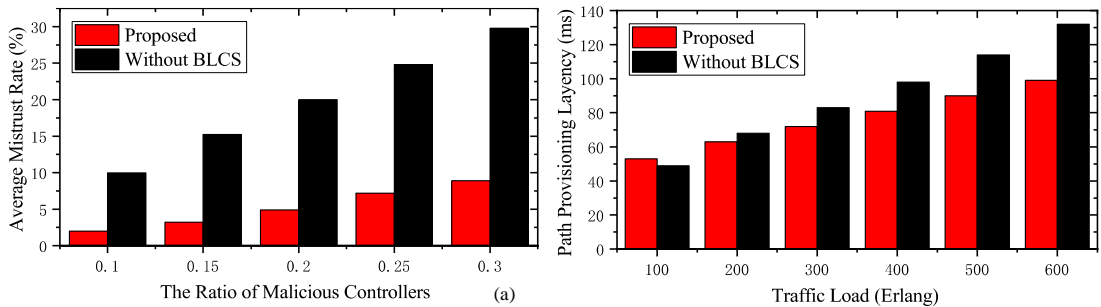

Fig. 4 (a) Average mistrust rate, (b) path provisioning latency, under different traffic load.

Impact on Packet Loss Probability: Fig. 5(a) shows the impact on packet loss probability with increasing traffic load in the CPS. It can be inferred from Fig. 5(a) that the packet loss probability for the proposed scheme is minimal compared to the scenario without the proposed scheme. The proposed scheme assigns the routes using the BLCS in such a way it decreases the packet loss probability, further improving the distributed control security, which proves the efficiency of our scheme.

Impact on Blocking Probability: Fig. 4(d) shows the variations of the overall blocking probability with respect to the traffic load. This figure shows that the proposed scheme also reduces the blocking probability compared to the scheme without BLCS. It can be inferred from Fig. 4(a), Fig. 4(b) and Fig. 5(b) that the proposed scheme selects trusted controllers as well as perform faster path provisioning during the requests processing, resulting in the reduction of blocking probability.

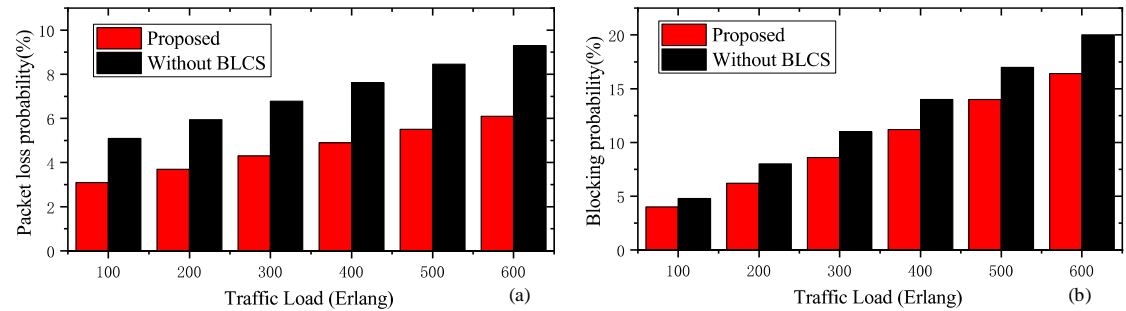

Fig. 5 (a) Packet loss probability and (b) blocking probability under different traffic load.

## VII. Conclusion and Future work

This article presents a novel brain-like based distributed control security architecture in fog radio and optical networks for CPS, which can solve the cyber security issue without the privacy exposure. The functional entities of the architecture and interworking procedure in secure control mode are presented and investigated. The performances are demonstrated on the testbed for secure distributed control. We also assess its performances in the scenario with heavy traffic load and compare it with the other scheme. Numerical results show that BLS scheme can locate malicious controllers and security routing, while reduce the average mistrust rate, path provisioning latency, packet loss probability and blocking probability. In the future, we will research more sophisticated optimization techniques along with unsupervised learning and reinforcement learning based BLCS. These can further enhance the security, reliability, and accuracy of rapidly growing CPS architectures.

## Acknowledgement

This work has been supported in part by NSFC project (61871056), Young Elite Scientists Sponsorship Program by CAST (YESS) (2018QNRC001), Fundamental Research Funds for the Central Universities (2018XKJC06, 2019PTB-009) and Open Fund of SKL of IPOC (BUPT) (IPOC2018A001), SKL of IPOC (BUPT).